\documentclass[letterpaper, 10 pt, conference]{ieeeconf}  

\IEEEoverridecommandlockouts                              

\title{\LARGE 
Implicit Trajectory Planning for Feedback Linearizable Systems: A Time-varying Optimization Approach}
\author{Tianqi Zheng, John Simpson-Porco, and Enrique Mallada\thanks{T. Zheng and E. Mallada are with the Department of Electrical and Computer Engineering, Johns Hopkins University, Baltimore, MD 21218, USA. Email: {\tt\small \{tzheng8, mallada\}@jhu.edu}. 
John Simpson-Porco is with the Department of Electrical and Computer Engineering, University of Waterloo, Waterloo ON, N2L 3G1 Canada. Email:{\tt\small jwsimpson@uwaterloo.ca}. 
This work was supported by ARO through contract W911NF-17-1-0092,  NSF through grants CNS 1544771, EPCN 1711188, AMPS 1736448, and CAREER 1752362, and Johns Hopkins University Discovery Award.
}}

\usepackage{amsthm}
\usepackage{subfiles}
\usepackage{amsmath}
\allowdisplaybreaks
\usepackage{amsfonts}
\usepackage{amssymb}
\usepackage{environ}
\usepackage{accents}
\usepackage{balance}
\usepackage{verbatim}
\usepackage{mathtools}
\usepackage{lineno}
\usepackage{graphicx}
\usepackage{subfig}
\usepackage{cite} 
 
\usepackage{ifthen}
    \newboolean{arxiv}
    \setboolean{arxiv}{true}

\newtheorem{thm}{Theorem}
\newtheorem{lem}[thm]{Lemma}

\newtheorem{defn}{Definition}

\newtheorem{prob}{Problem}

\newtheorem{ass}{Assumption}
\allowdisplaybreaks

\let\labelindent\relax 
\usepackage{enumitem}
\setlist[enumerate]{leftmargin=*}
\setlist[itemize]{leftmargin=*}
 
\usepackage{ifthen}
\newboolean{showcomments}
\setboolean{showcomments}{true}
\usepackage[usenames,dvipsnames]{xcolor}
\usepackage{todonotes}

\definecolor{bleudefrance}{rgb}{0.19, 0.55, 0.91}
\definecolor{ao(english)}{rgb}{0.0, 0.5, 0.0}

\newcommand{\addcite}[0]{\ifthenelse{\boolean{showcomments}}
{\textcolor{purple}{(add cite(s)) }}{}}%

\newcommand{\enrique}[1]{  \ifthenelse{\boolean{showcomments}}
{\todo[inline,color=bleudefrance]{Enrique: #1}}{}}
\newcommand{\emmargin}[1]{\ifthenelse{\boolean{showcomments}}{\marginpar{\color{bleudefrance}\tiny EM: #1}}{}}

\newboolean{showedits}
\setboolean{showedits}{true}
\usepackage[xcolor=bleudefrance]{changes}
\definechangesauthor[color=bleudefrance]{EM}
\newcommand{\aem}[1]{
\ifthenelse{\boolean{showedits}}
{\added[id=EM]{#1}}
{#1}
}
\newcommand{\chem}[2]{
\ifthenelse{\boolean{showedits}}
{\replaced[id=EM]{#1}{#2}}
{#1}
}
\newcommand{\dem}[1]{
\ifthenelse{\boolean{showedits}}
{\deleted[id=EM]{#1}}
{}
}

\newcommand{\real}[0]{\mathbb R}

\DeclareSymbolFont{bbold}{U}{bbold}{m}{n}
\DeclareSymbolFontAlphabet{\mathbbold}{bbold}

\DeclareMathOperator*{\spec}{spec}

\newcommand{\diag}[1]{\ensuremath{\mathrm{diag}\left(#1\right)}}

\DeclarePairedDelimiter\norm{\lVert}{\rVert}

\newcommand{\eucd}{\mathbb{R}}
\newcommand{\col}[2]{\left(#1,\dots,#2\right)}

\ifthenelse{\boolean{arxiv}}{}{
\usepackage{setspace}
\setstretch{.95}
}

\begin{document}

\maketitle
\begin{abstract}
    We develop an optimization-based framework for joint real-time trajectory planning and feedback control of feedback-linearizable systems. To achieve this goal, we define a target trajectory as the optimal solution of a time-varying optimization problem. In general, however, such trajectory may not be feasible due to, e.g., nonholonomic constraints. To solve this problem, we design a control law that generates feasible trajectories that asymptotically converge to the target trajectory. More precisely, for systems that are (dynamic) full-state linearizable, the proposed control law implicitly transforms the nonlinear system into an optimization algorithm of sufficiently high order. We prove global exponential convergence to the target trajectory for both the optimization algorithm and the original system. We illustrate the effectiveness of our proposed method on multi-object or multi-agent tracking problems with constraints.
\end{abstract}


\section{Introduction}

{The ability to design and execute safe trajectories for nonlinear systems constitutes one of the major pillars towards the development of autonomous systems \cite{chiang1988stability,encarnacao20003d,porfiri2007tracking,murray1993nonholonomic,werling2010optimal,mellinger2011minimum,fiorini1998motion,mellinger2012mixed}.} Thus, not surprisingly, motion planning and control has been an increasingly popular subject of research in both industry and academia \cite{encarnacao20003d,porfiri2007tracking,werling2010optimal,murray1993nonholonomic,mellinger2011minimum,paden2016survey,kalmar2004near,mellinger2012mixed,fiorini1998motion}. In general, this problem is usually solved in a two stage-approach. The first stage, known as motion planning, designs trajectories \textemdash{} usually by solving an optimization problem \textemdash{} that are \emph{feasible}, or in other words, trajectories that account for obstacles and system constraints \cite{werling2010optimal,mellinger2011minimum,fiorini1998motion,mellinger2012mixed}. In the second stage, feedback controllers are designed to track the designed trajectories and account for system uncertainties and disturbances \cite{encarnacao20003d,porfiri2007tracking,paden2016survey,mellinger2012mixed}.

While in general this approach has been quite successful, it requires the planning problem to be solved quickly enough to account for time varying environments. Thus, it imposes limits on the complexity of the optimization problem that implements motion planning. In particular, when implemented in real-time, motion planning usually amounts to linear~\cite{mellinger2011minimum} or quadratic optimization problems~\cite{kalmar2004near}, and rarely involves more than one agent at a time. In this work, we seek to alleviate these limitations by combining the planning and tracking stages.

More precisely, we seek to develop a time-varying optimization based framework for joint real-time trajectory planning and feedback control of a nonlinear dynamical system. To achieve this goal, we first define a target trajectory; i.e., a minimizing path, as the optimal solution of a time-varying optimization problem. Although in principle the target trajectory may not be feasible due to initial conditions or nonholonomic constraints, we overcome this problem by designing a control law that exponentially drives the system towards the target trajectory. For nonlinear systems that are dynamic full-state linearizable, we accomplish this by designing a control law that transforms the nonlinear system into an optimization algorithm.



Our work broadly aligns with the extensive research recently performed at the intersection of optimization and traditional control theory~\cite{helmke2012optimization,cherukuri2016asymptotic,fazlyab2017variational}, and more precisely with recent works trying to eliminate the time-scale separation usually present between optimization and control~\cite{nelson2018integral,colombino2019online,lawrence2018linear}. In many practical settings of robot control, especially when designing control laws for multi-robot system tracking of moving objects \cite{zhou2011multirobot,lee2015multirobot}, the optimization problems are not stationary (i.e., time-invariant), as the objective function and/or the constraints depend explicitly on time. 
Such time-varying optimization problems with or without constraints have been studied in both continuous \cite{fazlyab2017prediction} and discrete time settings \cite{simonetto2017prediction} using prediction-correction algorithms. {Our work here can be understood as an extension of these ideas to accommodate non-trivial system dynamics.}


\ifthenelse{\boolean{arxiv}}{ 
The rest of the paper is organized as follows. Section \ref{sec:preliminaries} introduces some preliminary definitions, including feedback linearization, which means a system can be transformed into a linear system by a state diffeomorphism, its dynamic feedback extension, and elementary analysis of Hurwitz linear systems. Then, in Section \ref{sec:ProblemStatement}, we formally state the problem and present two motivating example with different system dynamics (integrator and wheeled mobile robot). The main contribution of this paper is contained in Section \ref{sec:4}, where we use a prediction-correction algorithm for the time-varying optimization and feedback linearization to satisfy the design requirement. We design a control law which (i) implicitly defines a target trajectory as the optimal solution of a time-varying optimization problem, and (ii) asymptotically drives the system to the target trajectory. Finally, we illustrate the effectiveness of our approach in two examples, one where a wheeled mobile robot switches from tracking one moving object to another (Section \ref{sec:simulation:1}), and another where multiple agents must track multiple objects with internal distance constraints (Section \ref{sec:simulation:2}). 
}
{The rest of the paper is organized as follows. Section \ref{sec:preliminaries} introduces some preliminary definitions, including feedback linearization, which means a system can be transformed into a linear system by a state diffeomorphism and its dynamic feedback extension. Then, in Section \ref{sec:ProblemStatement}, we formally state the problem and present two motivating example with different system dynamics (integrator and wheeled mobile robot). The main contribution of this paper is contained in Section \ref{sec:4}, where we use a prediction-correction algorithm for the time-varying optimization and feedback linearization to satisfy the design requirement. We design a control law which (i) implicitly defines a target trajectory as the optimal solution of a time-varying optimization problem, and (ii) asymptotically drives the system to the target trajectory. Finally, we illustrate the effectiveness of our approach in two examples, one where a wheeled mobile robot switches from tracking one moving object to another (Section \ref{sec:simulation:1}), and another where multiple agents must track multiple objects with internal distance constraints (Section \ref{sec:simulation:2}). }

\subsubsection*{Notation} Given an $n$-tuple $(x_1, ...,x_n)$, $\mathbf{x} \in \eucd^n$ is the associated column vector. The $n \times n $ identity matrix is denoted as $ \mathbf{I}_n $. For a square symmetric matrix $\bf{A}$, is positive (semi-)definite, and write $\bf{A}\succ 0$ ($\bf{A} \succeq 0$), if and only if all the eigenvalues of $\bf{A}$ are positive (nonnegative). We further write $\bf{A}\succ \bf{B}$ ($\bf{A}\succeq \bf{B}$) whenever $\bf{A-B}\succ0$ ($\bf{A-B}\succeq0$). The Frobenius norm of a vector $\bf{x}$ is denoted by $\norm{\mathbf{x}}_2$, and the Euclidean norm of a matrix $\bf{A}$ by $\mathbf{\norm{A}}_2$.

Given a continuously differentiable function $f(\mathbf{x},t)$ of state $\mathbf{x} \in \mathbb{R}^n$ and time $t \in \real$, the gradient with respect to $\mathbf{x}$ (resp. $t$) is denoted by $\nabla_\mathbf{x} f(\mathbf{x},t)$  (resp. $\nabla_t f(\mathbf{x},t)$). The total derivative of $\nabla_\mathbf{x} f(\mathbf{x}(t),t)$ with respect to $t$ is denoted by $\dot{\nabla}_\mathbf{x} f(\mathbf{x},t):=\frac{d}{dt}{\nabla}_\mathbf{x} f(\mathbf{x}(t),t)$, and the
$n$-th total derivative with respect to $t$ by $\nabla^{(n)}_\mathbf{x} f(\mathbf{x},t)$.
The partial derivatives of $\nabla_\mathbf{x} f(\mathbf{x},t)$ with respect to $\mathbf{x}$ and $t$ are denoted by $\nabla_{\mathbf{xx}}f(\mathbf{x},t) := \frac{\partial}{\partial \mathbf{x}} \nabla_\mathbf{x} f(\mathbf{x},t) \in \eucd^{n\times n}$ and $\nabla _{\mathbf{x}t}f(\mathbf{x},t):=\frac{\partial}{\partial t}{\nabla}_\mathbf{x} f(\mathbf{x},t) \in \eucd ^{n}$ , respectively.  The derivative $L_f h$ of a function $h:\eucd^n\rightarrow\eucd$ along the vector field $ f :\eucd^n\rightarrow \eucd^n$ is given by $(L_fh)(\mathbf{x}) = \nabla h(\mathbf{x})^Tf(\mathbf{x})$. Taking the derivative of $h$ first along a vector field $f$ and then along a vector field $g$ is given by $(L_gL_fh)(\mathbf{x}) = \frac{\partial(L_fh)}{\partial x} g(\mathbf{x})$. If $h$ is being differentiated $k$ times along $f$, the notation $L^k_fh(\mathbf{x}) = \frac{\partial(L^{k-1}_fh)}{\partial x} g(\mathbf{x})$ is used.

\section{Preliminaries}\label{sec:preliminaries}

\subsection{Feedback Linearization}\label{ssec:feedback-lin}

\subsubsection{Static Feedback Linearization}
We consider a square control-affine nonlinear system with the state $\mathbf{x} \in D \subset \eucd^n$, $m$ inputs $\mathbf{u} \in \mathbb{R}^{m}$ and $m$ outputs $\mathbf{y} \in \mathbb{R}^{m}$, described in state-space form:
\begin{subequations}\label{eq:system}
    \begin{align}
      \Dot{\mathbf{x}} &= f(\mathbf{x}) + \mathbf{G(x)} \mathbf{u}\;,  \label{eq:system-state}\\
     \mathbf{y} &= h(\mathbf{x})\;, \label{eq:system-output}
\end{align}
\end{subequations}
where $f: D \to \eucd ^n$, $\mathbf{G}: D \to \eucd ^{n\times m}$, and $h: D \to \eucd^{m}$ are sufficiently smooth on a domain $D \subset \eucd ^n$, with $\mathbf{G}$ and $h$ expanded as 
\begin{align*}
    \mathbf{G}(\mathbf{x})  &= \big[g_1(\mathbf{x}), \dots,g_m(\mathbf{x})\big] \in \eucd^{n \times m},\\
    h(\mathbf{x}) & = \col{h_1(\mathbf{x})}{h_m(\mathbf{x})} \in \eucd^{m}.
\end{align*}



\begin{prob}[State-Space Exact Linearization]\label{Defn:State_Space_Exact_Linearization}
Given a point $\mathbf{x_0} \in D \subset \eucd^{n}$, for the control-affine nonlinear system \eqref{eq:system}, find a feedback controller $\mathbf{u} = \alpha(\mathbf{x})+\beta(\mathbf{x})\mathbf{v}$ defined on a neighborhood $U$ of $\mathbf{x_0}$, a coordinate transformation $\mathbf{z} = \Phi(\mathbf{x})$ also defined on $U$, and a controllable pair $(\mathbf{A,B})$ $(\mathbf{A} \in \eucd ^{n \times n}, \mathbf{B}\in \eucd ^{n \times m})$ such that: 
\begin{align*}
  \dot{\mathbf{z}}&\!=\! \mathbf{Az}\!+\!\mathbf{Bv}\!=\!\frac{   \partial\Phi(\mathbf{x})}{  \partial \mathbf{x}}\Big(f(\mathbf{x})\!+\!g(\mathbf{x})(\alpha(\mathbf{x})\!+\!\beta(\mathbf{x})\mathbf{v})\Big).
\end{align*}
\end{prob}

The key condition on \eqref{eq:system} for solvability of the State-Space Exact Linearization Problem is that the system possesses vector relative degree \cite{isidori2013nonlinear}. In other references \cite{sastry2013nonlinear}, this is also called \textit{Full State Linearization}.


\begin{defn}[Vector Relative Degree \cite{isidori2013nonlinear}]\label{def:relative-degree}
The control affine system \eqref{eq:system} is said to have \emph{vector relative degree} $\{r_1,r_2, \dots ,r_m \} $ at a point $\mathbf{x_0}\mathbf{ }\in  D \subset \eucd^{n}$ if: 
\begin{enumerate}[label=(\roman*),leftmargin=*]\label{R(x)}
    \item $L_{g_j}L^k_f h_i(\mathbf{x}) = 0$ for all $1 \leq i \leq m$, for all $k < r_i -1$, for all $1 \leq j \leq m$, and for all $\mathbf{x}$ in a neighborhood of $\mathbf{x_0}$, and
    \item the $m \times m$ matrix,
\begin{align}
    \mathbf{R(x)} \!=\! 
    \begin{bmatrix*}
L_{g_1}L^{r_1 -1}_f h_1(\mathbf{x})\!&\! \dots \!&\! L_{g_m}L^{r_1 -1}_f h_1(\mathbf{x})\\
L_{g_1}L^{r_2 -1}_f h_2(\mathbf{x})\!&\! \dots \!&\! L_{g_m}L^{r_2 -1}_f h_2(\mathbf{x})\\
\vdots \!&\! \dots \!&\! \vdots \\
L_{g_1}L^{r_m -1}_f h_m(\mathbf{x})\!&\! \dots \!&\! L_{g_m}L^{r_m -1}_f h_m(\mathbf{x})
    \end{bmatrix*},\label{eq:relative-degree-matrix}
\end{align}
is nonsingular at $\mathbf{x=x_0}$.
\end{enumerate}
\end{defn}

\begin{lem}[Solution of Exact Linearization Problem with static feedback linearization {\cite[Lemma 5.2.1]{isidori2013nonlinear}}]\label{Lem:State_Space_Exact_linearization}
Suppose the matrix $\mathbf{G}(\mathbf{x_0})$ has rank $m$. Then the State-Space Exact Linearization Problem is solvable if and only if there exists a neighborhood of $\mathbf{x_0}$ such that the system \eqref{eq:system} has vector relative degree $ \{r_1,r_2, \dots ,r_m \}$ at $\mathbf{x_0}$ and $r_1+r_2+\dots+r_m=n$. In particular, one may choose
\begin{enumerate}
\item[(i)] the feedback as
\[
\mathbf{u} = -\mathbf{R(x)}^{-1}\mathbf{p(x)} + \mathbf{R(x)}^{-1}\mathbf{v},
\]
where $\mathbf{p(x)} = \mathrm{col}(L^{r_1}_{f} h_1(\mathbf{x}),   \dots , L^{r_m}_{f} h_m(\mathbf{x})) \in \eucd^m$ and $\mathbf{R(x)}$ is defined in  \eqref{eq:relative-degree-matrix},
\item[(ii)] the coordinate transformation as
\[
\Phi(\mathbf{x}) = \mathrm{col}(h_1(\mathbf{x}),\dots,L_f^{r_1-1}h_1(\mathbf{x}),\dots,L_f^{r_m-1}h_m(\mathbf{x})),
\]
\item[(iii)] $\mathbf{(A,B)}$ having the \emph{Brunovsky Canonical Form} 
\begin{align*}
    \mathbf{A} = \diag{\mathbf{A}_1,\dots,\mathbf{A}_m}, \; \mathbf{B }= \diag{\mathbf{b}_1,\dots,\mathbf{b}_m},
\end{align*}
where $\mathbf{A}_i \in \eucd^{r_i \times r_i}$ and $\mathbf{b}_i \in \eucd^{r_i}$ are
\begin{align*}
    \mathbf{A}_i = \begin{bmatrix*}
    0&1&0&\dots& 0 \\
    0&0&1&\dots&0 \\
    .&.&.&\dots&.\\
    0&0&0&\dots&1\\
    0&0&0&\dots&0
    \end{bmatrix*}, \quad \mathbf{b}_i = \begin{bmatrix}0 \\ 0 \\ \vdots \\ 0 \\ 1\end{bmatrix}.
\end{align*}
\end{enumerate}

Remark: The input $v_i$ controls only the output $y_i$ throughout a chain of $r_i$ integrator. When $r_1+r_2+\dots+r_m=n$, in the closed loop system there are no unobservable dynamics.

\end{lem}

\subsubsection{Dynamic Feedback Linearization}

For systems which do not have vector relative degree, one can sometimes achieve a vector relative degree by introducing auxiliary state variables $ \boldsymbol{\zeta} $, e.g., for a system that is differentially flat \cite{murray1995differential}, by using dynamic feedback of the form
\begin{subequations}\label{eq:dynamic-feedback-linearization}
\begin{align}
 &  \mathbf{u} = \alpha(\mathbf{x},\boldsymbol{\zeta}) + \beta(\mathbf{x},\boldsymbol{ \zeta})\mathbf{w} ,\\
 &   \Dot{\boldsymbol{\zeta}} = \gamma(\mathbf{x},\boldsymbol{\zeta}) + \delta(\mathbf{x},\boldsymbol{ \zeta})\mathbf{w}. 
\end{align}
\end{subequations}
Consider then the composite system formed by \eqref{eq:system} and \eqref{eq:dynamic-feedback-linearization}
    \begin{align}
    \begin{bmatrix*}
    \dot{\mathbf{x}} \\ \Dot{\boldsymbol{\zeta}}
    \end{bmatrix*} = \Tilde{f}(\mathbf{x},\boldsymbol{\zeta})  +\Tilde{\mathbf{G}}(\mathbf{x},\boldsymbol{\zeta}) \mathbf{w}, \;\;
    \mathbf{y} = h(\mathbf{x}), \label{eq:composite-system}
    \end{align} 
where
\begin{align*}
    \Tilde{f}(\mathbf{x},\boldsymbol{\zeta}) \!= \!
    \begin{bmatrix*} f(\mathbf{x})\!+ \!\mathbf{G}(\mathbf{x})\alpha(\mathbf{x},\boldsymbol{\zeta}) \\ \gamma(\mathbf{x},\boldsymbol{\zeta})
    \end{bmatrix*}, \Tilde{\mathbf{G}}(\mathbf{x},\boldsymbol{\zeta}) \!= \!
    \begin{bmatrix*} g(\mathbf{x})\beta(\mathbf{x}, \boldsymbol{\zeta}) \\\delta(\mathbf{x}, \boldsymbol{\zeta})
    \end{bmatrix*}.
\end{align*}

The following is a direct extension of Lemma \ref{Lem:State_Space_Exact_linearization}. Further details on this approach, known as \textit{dynamic extension}, can be found in \cite{isidori2013nonlinear} and \cite{sastry2013nonlinear}. 

\begin{lem}[Solution of Exact Linearization Problem using dynamic feedback linearization {\cite{isidori2013nonlinear}}]\label{Lem:Dynamic_Feedback_Linearization}
Suppose the matrix $\Tilde{\mathbf{G}}(\mathbf{x_0},\boldsymbol{\zeta_0})$ has rank $m$. Then the State-Space Exact Linearization Problem is solvable if and only if there exists a neighborhood of $[\mathbf{x_0},\boldsymbol{\zeta_0}]^T$ such that the system \eqref{eq:composite-system} has vector relative degree $ \{r_1,r_2, \dots ,r_m \}$ at $[\mathbf{x_0},\boldsymbol{\zeta_0}]^T$ and $r_1+r_2+\dots+r_m=n$. In particular, one may choose
\begin{enumerate}
\item[(i)] 
The dynamic feedback defined by \eqref{eq:dynamic-feedback-linearization} and
\begin{align}
    \mathbf {w = -R^{-1}(x,\boldsymbol{\zeta})p(x,\boldsymbol{\zeta})+ R^{-1}(x,\boldsymbol{\zeta})v},\label{eq:Dynamic_Feedback_Function}
\end{align} where $ \mathbf{p}(\mathbf{x},\boldsymbol{\zeta})= \mathrm{col}( L^{r_1}_{\Tilde{f}} h_1(\mathbf{x}),\dots, L^{r_m}_{\Tilde{f}} h_m(\mathbf{x}))\in \eucd^m$ and $\mathbf{R(x},\boldsymbol{\zeta})$ is defined in \eqref{eq:relative-degree-matrix}.
\item[(ii)] the coordinate transformation as
\[
\Phi(\mathbf{x},\boldsymbol{\zeta}) \!=\! \mathrm{col}(h_1(\mathbf{x}),\!\dots\!,L_{\tilde{f}}^{r_1-1}h_1(\mathbf{x}),\!\dots\!,L_{\tilde{f}}^{r_m-1}h_m(\mathbf{x})),
\]
\item[(iii)] $\mathbf{(A,B)}$ having the \emph{Brunovsky Canonical Form}.
\end{enumerate}
\end{lem}


\ifthenelse{\boolean{arxiv}}{ 

\subsection{Convergence Rate of Hurwitz Matrix}
A square matrix $\mathbf{H}$ is called Hurwitz if
\begin{align*}
    \mu (\mathbf{H}) := \max_{\lambda\in\spec(\mathbf{H})} {\Re[\lambda]} <0\;,
\end{align*}
where $\spec(\mathbf{H}):=\{\lambda_i\}$ denotes the set of eigenvalues of $\mathbf{H}$. If $\mathbf{H}$ is Hurwitz, then $\lim _{t \to +\infty} e^{\mathbf{H}t} = 0$. 


\begin{thm}[Exponential Convergence of Hurwitz Matrices~{{\cite[Theorem 8.1]{hespanha2018linear}}}]\label{thm:exp-convergence}
If $\mathbf{H}$ is Hurwitz, then there exist constants $c, \alpha > 0$ such that
\begin{align*}
   & \norm{e^{\mathbf{\mathbf{H}}t}}_2 \leq c e^{-\alpha t}, \quad \text{for all} \,\,\, t \geq 0,
\end{align*}
where  $-\alpha := \max_{\lambda\in\spec(\mathbf{H})} \Re[\lambda]+ \epsilon $, for some  $ \epsilon >0$ that are small enough.
\end{thm}

When $\mathbf{H}$ is diagonalizable, i.e., when all Jordan blocks of $\mathbf{H}$ have size equal to 1, one can choose $-\alpha =\max_{\lambda\in\spec(\mathbf{H})} \Re[\lambda]$.
}
{}

\section{Problem Statement}\label{sec:ProblemStatement}
As mentioned before, our goal is to develop an optimization based framework for joint real-time trajectory planning and feedback control of nonlinear systems. To achieve this goal we develop a two-stage design approach where we {(i)} implicitly define the desired trajectory as the optimal solution of a \emph{time-varying} optimization problem, and (ii) design a control law that seeks to converge asymptotically to the optimal solution of the optimization problem.

Formally, we consider a nonlinear system as described in \eqref{eq:system}. Let $t \geq  0$ be a continuous time index, and $f_0: \eucd ^m \times \eucd _+ \to \eucd$ 
be a time-varying function of the output $\mathbf{y}$. Using $f_0(\mathbf{y},t)$ we implicitly define our target trajectory; i.e., the \emph{minimizing path}:
\begin{equation}
    \mathbf{y}^*(t) = \arg\min_{\mathbf{y} \in \eucd^m} \; f_0(\mathbf{y},t). \label{eq:Time_Varing_Optimization_Problem}
\end{equation}

The goal is to generate a control input $\mathbf{u}(t)$ such that $\norm{ \mathbf{y}(t)- \mathbf{y}^*(t)}_2 \to 0$ as $t\to \infty$ for all initial conditions; i.e., global asymptotic convergence. The following assumption will be used throughout this paper.

\begin{ass}[Objective Function]\label{ass:convexity}
The objective function $f_0(\mathbf{y},t)$ is infinitely differentiable $( \mathbb{C}^\infty)$ with respect to both $\mathbf{y}$ and $t$, and is uniformly strongly convex in $\mathbf{y}$; i.e., $\nabla_{\mathbf{yy}} f_0(\mathbf{y}(t),t) \succeq m_f\mathbf{I}_m $ for all $t \geq 0$ and for some $m_f > 0$.
\end{ass}

The remainder of this section provides two examples that help motivate both our goals and our solution approach.
\subsection{Example \#1: Integrator}


We aim to design a control law for an integrator
\begin{equation}
    \begin{aligned}
        \mathbf{\dot{x} = u},\;\;
        \mathbf{y = x},
    \end{aligned}\label{eq:Integrator_System_Dynamic}
\end{equation}such that $\mathbf{y}$ converges asymptotically to the optimal solution of  time-varying optimization problem \eqref{eq:Time_Varing_Optimization_Problem}:
\begin{equation*}
    \mathbf{y}^*(t) = \arg\min_{\mathbf{y} } \; f_0(\mathbf{y},t). 
\end{equation*}

\begin{figure}[h]
\centering
\includegraphics[width=0.45\textwidth]{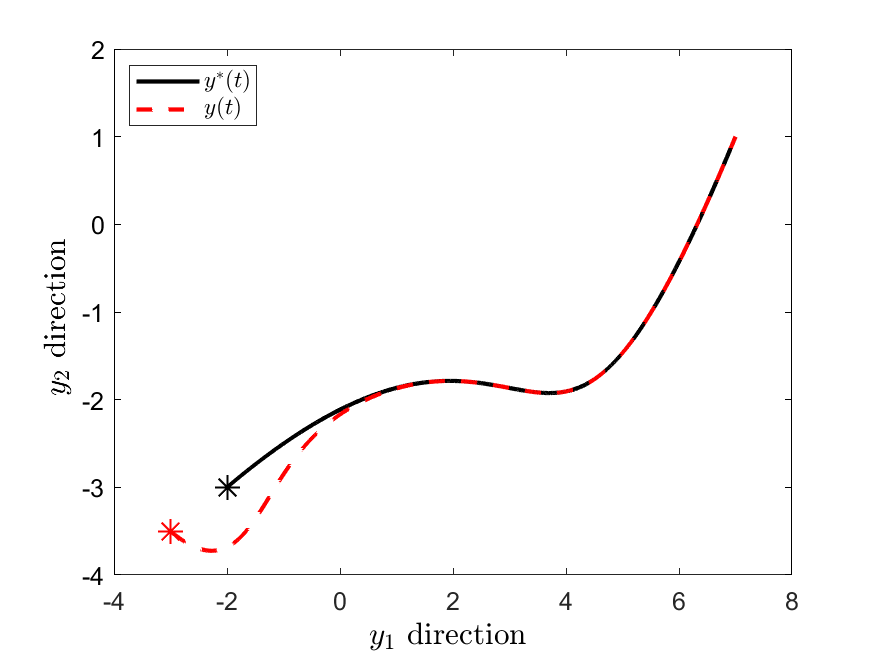}
    \caption{Plot of a robot tracking an object, where $\mathbf{y}^*(t)$ \eqref{eq:Time_Varing_Optimization_Problem} is simply the object trajectory and $\mathbf{y}(t)$ represents the real trajectory of robot. Due to miss matching initial conditions (highlighted using asterisk), we design a control law that converge asymptotically to the target trajectory $\mathbf{y}^*(t)$.}
    \label{fig:1}
    \end{figure}

Notice that, even though we can instantaneously change the speed and direction of $y(t)$ in \eqref{eq:Integrator_System_Dynamic}, the initial condition $y(0)$ may not match $y^*(0)$. This is illustrated in Figure \ref{fig:1}.
    
This problem can be overcome by finding a control law that transforms \eqref{eq:Integrator_System_Dynamic} into the following optimization dynamics
\begin{equation}
\dot \nabla_\mathbf{y} f_0(\mathbf{y},t) = -\mathbf{P}\nabla_\mathbf{y} f_0(\mathbf{y},t), \quad \mathbf{P}\succ 0,  \label{eq:Gradient-PositiveDefinite-System}
\end{equation}
where the gradient $\nabla_\mathbf{y} f_0(\mathbf{y},t)$ is driven to zero exponentially fast \cite{fazlyab2017prediction,simonetto2017prediction}.
Thus, since by convexity (see Assumption \ref{ass:convexity}), the optimal trajectory $\mathbf{y}^*(t)$ is characterized by $\nabla_\mathbf{y} f_0(\mathbf{y}^*(t),t) = 0$, the controlled $\mathbf{y}$ asymptotically reaches $\mathbf{y}^*(t)$. 

To achieve this transformation, we first characterize the required evolution of $\mathbf{y}$ for  \eqref{eq:Gradient-PositiveDefinite-System} to hold, and then define the proper control law.
Using the chain rule to differentiate the gradient term with respect to time yields
\begin{align*}
    \dot{\nabla}_\mathbf{y} f_0(\mathbf{y},t) = \nabla_{\mathbf{yy}}f_0(\mathbf{y},t)\dot{\mathbf{y}}+ \nabla_{\mathbf{y}t}f_0(\mathbf{y},t).
\end{align*}
Then, by  combining \eqref{eq:Gradient-PositiveDefinite-System} and the above equation, we find that $\mathbf{\dot y}$ is implicitly defined by
\begin{align}
   \mathbf{ \dot y}_{\rm imp} = -\nabla_{\mathbf{yy}}^{-1}f_0(\mathbf{y},t)[\mathbf{P}\nabla_\mathbf{y} f_0(\mathbf{y},t)
    +\nabla_{\mathbf{y}t}f_0(\mathbf{y},t)]. \label{eqn:integrator-implicit-trajectory}
\end{align}
Finally, since by \eqref{eq:Integrator_System_Dynamic}, $\mathbf{u=\dot y}$, equation \eqref{eqn:integrator-implicit-trajectory} leads to the control:
\begin{align*}
    \mathbf{u}= -\nabla_{\mathbf{yy}}^{-1}f_0(\mathbf{y},t)[\mathbf{P}\nabla_\mathbf{y} f_0(\mathbf{y},t)
    +\nabla_{\mathbf{y}t}f_0(\mathbf{y},t)]. 
\end{align*}

The above control law implicitly transforms \eqref{eq:Integrator_System_Dynamic} into \eqref{eq:Gradient-PositiveDefinite-System}. Further, it has a nice optimization-based interpretation consisting of two terms~ \cite{fazlyab2017prediction,simonetto2017prediction}:
\begin{enumerate}
    \item a \textit{prediction term} $-\nabla_{\mathbf{yy}}^{-1}f_0(\mathbf{y},t)\nabla_{\mathbf{y}t}f_0(\mathbf{y},t),$which tracks the change of the optimal solution; i.e., target trajectory,
    \item and a \textit{correction term} $-\nabla_{\mathbf{yy}}^{-1}f_0(\mathbf{y},t)\mathbf{P}\nabla_\mathbf{y} f_0(\mathbf{y},t),$ which acts as a proportional controller that cancels the optimality error and drives the system toward the optimum.
\end{enumerate}

Unfortunately, the solution approach shown in this example critically relies on the integrator structure in \eqref{eq:Integrator_System_Dynamic} that allows one to arbitrarily control $\mathbf{\dot y}$ by choosing $\mathbf{u}$.  However, for a general nonlinear system, satisfying \eqref{eq:Gradient-PositiveDefinite-System} may not be possible. This is shown in the next example.


\subsection{Example \#2: Wheeled Mobile Robot}\label{subsec:3B}
We now show how to extend the approach described above for a more involved example where we aim to drive a nonholonomic wheeled mobile robot (WMR)
\cite{oriolo2002wmr,sastry2013nonlinear}:
\begin{subequations}\label{eq:Unicycle_System_Dynamic}
\begin{align}
&\Dot{x}_1 =\cos(x_3)u_1,\\
&\Dot{x}_2=\sin(x_3)u_1,\\ 
&\Dot{x}_3 =u_2,\\
&\mathbf{y} = (x_1,x_2),
\end{align}
\end{subequations}
such that $\mathbf{y}$ converges asymptotically to the optimal solution of time-varying optimization problem \eqref{eq:Time_Varing_Optimization_Problem}.
If we once again want \eqref{eq:Unicycle_System_Dynamic} to match the dynamics \eqref{eq:Gradient-PositiveDefinite-System}, we need \eqref{eqn:integrator-implicit-trajectory} to hold.
However, it follows from \eqref{eq:Unicycle_System_Dynamic} that $\mathbf{\dot y} =[\cos(x_3)u_1,\sin(x_3)u_1]^T$, which is ill-defined.
%
It is obvious that one cannot control every direction of $\mathbf{\dot y}$ with this ill-defined equation therefore cannot derive a control law that ensures \eqref{eqn:integrator-implicit-trajectory}. 

This motivates the search for an alternative to \eqref{eq:Gradient-PositiveDefinite-System} that has the equivalent effect of driving $\mathbf{y}$ towards $\mathbf{y}^*(t)$. Instead, we seek to transform \eqref{eq:Unicycle_System_Dynamic} into
\begin{align}
    \begin{bmatrix*}
    \dot{ \nabla}_\mathbf{y} f_0(\mathbf{y},t) \\ \ddot{ \nabla}_\mathbf{y} f_0(\mathbf{y},t)
    \end{bmatrix*} = \begin{bmatrix*}
     0 & \mathbf{I}_m \\ -k_{\rm p} \mathbf{I}_m  &-k_{\rm d} \mathbf{I}_m
      \end{bmatrix*} \begin{bmatrix}
       \nabla_\mathbf{y} f_0(\mathbf{y},t) \\ \dot{\nabla}_\mathbf{y} f_0(\mathbf{y},t)
    \end{bmatrix},\label{eq:Second_order_Optimality_Error} 
\end{align}
where $k_{\rm p}, k_{\rm d} > 0$, which defines a Hurwitz matrix, and $ \mathrm{col}(\nabla_\mathbf{y} f_0(\mathbf{y},t),\dot{\nabla}_\mathbf{y} f_0(\mathbf{y},t)) $ can be interpreted as the optimality error of $\mathbf{y}$, and its time derivative. 

To find the control law that transforms \eqref{eq:Unicycle_System_Dynamic} into \eqref{eq:Second_order_Optimality_Error}, we can differentiate the gradient term with respect to time twice:
\begin{align*}
     \ddot{\nabla}_{\mathbf{y}}f_0(\mathbf{y},t) = &\nabla_{\mathbf{yy}}f_0(\mathbf{y},t)\mathbf{\ddot{y}}+ \dot{\nabla}_{\mathbf{yy}}f_0(\mathbf{y},t)\mathbf{\dot{y}}+\dot{ \nabla}_{\mathbf{y}t}f_0(\mathbf{y},t). 
\end{align*} 

Now combining once again the second row of \eqref{eq:Second_order_Optimality_Error} and the above equation leads to the following implicit condition for the acceleration $\mathbf{\ddot y}$:
\begin{align}\label{eq:ddot-y+grad}
   \mathbf{ \ddot{y}}_{\rm imp} 
    = -\nabla_{\mathbf{yy}}^{-1}f_0(\mathbf{y},t)\big[\dot{\nabla}_{\mathbf{yy}}f_0(\mathbf{y},t)\mathbf{\dot{y}}+\dot{ \nabla}_{\mathbf{y}t}f_0(\mathbf{y},t) \nonumber\\
    +k_{\rm p}\nabla_\mathbf{y} f_0(\mathbf{y},t)+k_{\rm d}\dot{\nabla}_\mathbf{y} f_0(\mathbf{y},t)\big]
\end{align}

Finally, by differentiating $\mathbf{y}$ with respect to time twice we notice that
the matrix on the right-hand side of
\begin{equation}\label{eq:Unicycle_yddot}
\mathbf{\ddot y }= \begin{bmatrix*}
     \cos(x_3)  & -\sin(x_3)u_1\\
     \sin(x_3)  & \cos(x_3)u_1 
    \end{bmatrix*}
    \begin{bmatrix*}
    \dot{u}_1 \\ u_2
    \end{bmatrix*}
\end{equation}
is invertible for every nonzero $u_1$ and thus, we can use $(\dot u_1,u_2)$ to control $\mathbf{\ddot y}$ to follow \eqref{eq:ddot-y+grad}:
\begin{align*}
    \begin{bmatrix*}
    \dot{u}_1 \\ u_2
    \end{bmatrix*} = \begin{bmatrix*}
     \cos(x_3)  & -\sin(x_3)u_1\\
     \sin(x_3)  & \cos(x_3)u_1 
    \end{bmatrix*}^{-1} \mathbf{\ddot{y}}_{\rm imp}.\label{eq:second_order_control_law}
\end{align*}As long as $u_1 \neq 0$, the control law is well-defined by introducing $u_1$ as an auxiliary state.

We finalize this section showing a particular case of \eqref{eq:ddot-y+grad} that is familiar for most control audience. If the task is simply tracking a moving object, we can define the following time-varying problem:
\begin{equation*}
 \mathbf{y}^*(t) =\arg\min_{\mathbf{y}} \tfrac{1}{2} \|\mathbf{y} - \mathbf{y}_{\rm d}(t)\|^2_2,
\end{equation*}
where $\mathbf{y}_{\rm d}(t)$ represents the target trajectory. And according to \eqref{eq:ddot-y+grad}, the implicitly defined trajectory takes the form:
\begin{align*}
    \mathbf{\ddot y}_{\rm imp} = & \;
    \mathbf{\ddot{y}}_{\rm d}(t)
    - k_p(\mathbf{y}-\mathbf{y}_{\rm d}(t))  - k_d(\mathbf{\dot{y}}-\mathbf{\dot{y}}_{\rm d}(t)). 
\end{align*}
Thus, in this case equation $\mathbf{\ddot y}_{\rm imp}$ can be interpreted as a common Proportional-Derivative (PD) controller.

\section{Implicit Trajectory Planning for Feedback Linearizable Systems}\label{sec:4}
The above motivating example shows how to extend the algorithm from a first-order system (an integrator) to a second-order system (a unicycle). In this Section, we aim to carry this procedure over to a more general setting.
 \ifthenelse{\boolean{arxiv}}{ 
In section \ref{subsec:4A}, we begin with a relatively restrictive assumption, where all $m$ output channels have same relative degree. This extension comes naturally from the WMR example in \ref{subsec:3B} by considering higher orders of the gradient as generalized optimality error. Then in Section \ref{sub:4B}, we relax the assumption so that the relative degree of each channel are not necessarily equal.}
{}

\ifthenelse{\boolean{arxiv}}{ 
\subsection{Uniform Vector Relative Degree}\label{subsec:4A}

We assume now that the system under consideration has a uniform vector relative degree, which will in general need to be achieved via dynamic extension. This is a natural extension from the WMR model, where the vector relative degree is $\{2,2\}$ and $n=4$. 

\begin{ass}[Uniform Vector Relative Degree]\label{ass:Uniform_Vector_relative_Degree}
The multivariable nonlinear system \eqref{eq:composite-system} has vector relative degree $r_1=\dots=r_m=k$ and $m\times k=n$.
\end{ass}



Based on Lemma \ref{Lem:Dynamic_Feedback_Linearization}, it is straightforward that for a multivariable nonlinear system satisfying Assumption \ref{ass:Uniform_Vector_relative_Degree}, the feedback function \eqref{eq:Dynamic_Feedback_Function} and a state diffeomorphism $ \mathbf{z}= \Phi(\mathbf{x},\boldsymbol{\zeta})$ will transform the composite system \eqref{eq:composite-system} into $\mathbf{\dot{z} = Az+Bv}$, with $(\mathbf{A,B})$ in Brunovsky Canonical Form. By computing the higher derivatives of output channel, we can implicitly design the trajectory for $\mathbf{y}$ using $\mathrm{col}(\nabla_\mathbf{y} f_0(\mathbf{y},t),\dots,\nabla^{(k-1)}_\mathbf{y} f_0(\mathbf{y},t))$ as a proxy for optimality error, where the goal is to construct the following dynamical system:
\begin{align}
    \begin{bmatrix}
       \dot{\nabla}_\mathbf{y} f_0(\mathbf{y},t) \\ \vdots \\ \nabla^{(k)}_\mathbf{y} f_0(\mathbf{y},t)
    \end{bmatrix} = \mathbf{H}
    \begin{bmatrix}
       \nabla_\mathbf{y} f_0(\mathbf{y},t) \\ \vdots \\ \nabla^{(k-1)}_\mathbf{y} f_0(\mathbf{y},t)
    \end{bmatrix}   ,\label{eq:K-th_order_Optimality_Error}
\end{align} with
\begin{align}
    \mathbf{H} = \begin{bmatrix*}
    0 &1 &0 &\dots &0 \\
    0 &0 &1 &\dots &0 \\
    \vdots & \vdots &  &\ddots & \vdots\\
    a_0 & a_1& a_2 &\dots &a_{k-1}
    \end{bmatrix*} \otimes \mathbf{I}_m \label{eq:Hurwitz_Matrix_Big_H}
\end{align}
being Hurwitz. The following technical lemma will be used during the calculation of new optimality error state.
\begin{lem}[Gradient Time Differentiation]\label{lem:Diffrentiating_Gradient_K_Times}
Differentiating the gradient $\nabla_\mathbf{y} f_0(\mathbf{y},t)$ with respect to time $k-$times yields:
\begin{equation}
\begin{aligned}
    \nabla_\mathbf{y}^{(k)} f_0(\mathbf{y},t) &= \sum_{m = 0}^{k-1} \binom{k-1}{m} \nabla_{\mathbf{yy}}^{(m)}f_0(\mathbf{y},t)\mathbf{y}^{(k-m)}\\
    &+\nabla_{\mathbf{y}t}^{(k-1)}f_0(\mathbf{y},t),
\end{aligned} \label{eq:Differentiate_gradient_K_Times}
\end{equation}
where $\binom{k-1}{m}$ represents the binomial coefficient.

\textit{Proof}: See Appendix  \ref{Prf:Diffrentiating_Gradient_K_Times}.\\
\end{lem}

Combining \eqref{eq:K-th_order_Optimality_Error} and \eqref{eq:Differentiate_gradient_K_Times}, we can implicitly design the trajectory for $\mathbf{y}$ by:
\begin{align}
& \mathbf{y}_{\rm imp}^{(k)} = \nabla_{\mathbf{yy}}^{-1}f_0(\mathbf{y},t)[\sum_{i = 0} ^{k-1} a_i\nabla_{\mathbf{y}}^{(i)}f_0(\mathbf{y},t)  \nonumber \\
& - \sum_{m=1}^{k-1}\binom{k-1}{m} \nabla_{\mathbf{yy}}^{(m)}f_0(\mathbf{y},t)\mathbf{y}^{(k-m)} -  \nabla_{\mathbf{y}t}^{(k-1)}f_0(\mathbf{y},t) ] \label{eq:y_kth_derivative_solution}.
\end{align}
Now, we formally provide our solution for systems with uniform relative degree.
\begin{thm}[Control Law for Uniform Vector Relative Degree Systems]\label{thm:Control_Law_uniform_Vector_Relative_Degree}
Consider the multivariable system defined as \eqref{eq:system} and the time-varying optimization problem defined as \eqref{eq:Time_Varing_Optimization_Problem}. If both assumptions \ref{ass:convexity} and \ref{ass:Uniform_Vector_relative_Degree} are satisfied, then the system will globally exponentially converge to the optimal solution of \eqref{eq:Time_Varing_Optimization_Problem}, by using  the control law:
\begin{align}
    \mathbf{u} = \alpha(\mathbf{x}, \boldsymbol{\zeta}) + \beta(\mathbf{x}, \boldsymbol{\zeta})\mathbf{R}(\mathbf{x},\boldsymbol{\zeta})^{-1}[\mathbf{y}_{\rm imp}^{(k)} -\mathbf{p}(\mathbf{x},\boldsymbol{\zeta}) ] \label{eq:final_Control_Input},
\end{align} where $\mathbf{y}_{\rm imp}^{(k)}$ is given in \eqref{eq:y_kth_derivative_solution} and the dynamic feedback function defined in \eqref{eq:Dynamic_Feedback_Function}. 
Moreover, the following inequalities hold:
\begin{align*}
    & \| \mathbf{y}(t)-\mathbf{y}^*(t)\|_2 \leq Ce^{-\alpha t} ,\\
    & 0 \leq f_0(\mathbf{y}(t),t) - f_0(\mathbf{y}^*(t),t) \leq m_fC^2e^{-2\alpha t},\\
    & 0 < C = \left(\tfrac{c^2}{m_f^2} \sum_{j=0}^{k-1}\nolimits \norm{{\nabla_\mathbf{y}^{(j)} f_0(\mathbf{y}(0),0)}}_2^2)\right)^{\frac{1}{2}} < \infty,\nonumber 
\end{align*}
for some constant $C > 0$, $  -\alpha = \max \{\Re(\lambda_i)+ \epsilon ,i \in [1...n]$\}, for some  $ \epsilon >0$ small enough.\\

\textit{Proof}: See Appendix \ref{prf:BCF_UniformRelativeDegree}.
\end{thm}

Theorem \ref{thm:Control_Law_uniform_Vector_Relative_Degree} makes a strong assumption on the structure of the nonlinear system, which is that the system must have equal vector degree $\{r_1= \dots =r_m \}$. In the next section we relax this assumption.
}
{}

\ifthenelse{\boolean{arxiv}}{ 
\subsection{Non-Uniform Vector Relative Degree}\label{sub:4B}

We now consider the less restrictive assumption.
\begin{ass}[Non-Uniform Vector Relative Degree]\label{ass:non-Unifrom_Vector_Relative_Degree}
The multivariable nonlinear system \eqref{eq:composite-system} has vector relative degree $\{r_1,\dots,r_m\}$ and $r_1+r_2+\dots+r_m=n$.
\end{ass}
As a result of Assumption \ref{ass:non-Unifrom_Vector_Relative_Degree}, the order of Lie differentiation of each channel is different (c.f.\eqref{eq:relative-degree-matrix}) and we cannot directly design the trajectory as in \eqref{eq:y_kth_derivative_solution}. However, remember that according to Lemma \ref{Lem:Dynamic_Feedback_Linearization}, the system is transformed into $\mathbf{\dot{z} = Az+Bv}$, with $(\mathbf{A,B})$ in Brunovsky Canonical Form. As a matter of fact, the input $v_i$ controls only the output $y_i$ throughout out a chain of $r_1$ integrators. If $\{r_1,...r_m\}$ are not equal, we can always introduce $k-r_i$ auxiliary states (integrators) for each channel $y_i$, where $k = \max \{ r_1,r_2, \dots ,r_m  \}$ and define the new input $s_i$ accordingly. Notice that this construction makes a dynamic extension of $\mathbf{\dot{z} = Az+Bv}$ that posses uniform order of Lie differentiation of each channel. For example, for channel $y_i$, we introduce the following states $
    \xi_1^i = v_i ,
    \xi_2^i = \dot{\xi}_1^i , 
    \dots,
    \dot{\xi}_{k-r_i}^i = s_i.$
 More specifically, the auxiliary states $\xi$ should satisfy the following dynamic:
\begin{subequations}\label{eq:unequal_auxilliary_state_xi_dyanmic_equation}
\begin{align}
    & \boldsymbol{ v} = \tilde{\alpha}(\boldsymbol{\xi}) + \tilde{\beta}(\boldsymbol{\xi})\boldsymbol{s},  \\ 
 &   \Dot{\boldsymbol{\xi}} =\tilde{ \gamma}(\boldsymbol{\xi}) +\tilde{ \delta}(\boldsymbol{\xi})\boldsymbol{s}.
\end{align}
\end{subequations}
Then the feedback function \eqref{eq:Dynamic_Feedback_Function}, the auxiliary states dynamic of $\boldsymbol{\xi}$ \eqref{eq:unequal_auxilliary_state_xi_dyanmic_equation}, and a state diffeomorphism $ \mathbf{z}= \Phi(\mathbf{x},\boldsymbol{\zeta,\xi})$ will transform the composite system \eqref{eq:composite-system} into $$\dot{\mathbf{z}} = \mathbf{Az+Bs},$$ with $\mathbf{A,B}$ in Brunovsky Canonical Form. 




\begin{thm}[Control Law for General Vector Relative Degree System]\label{thm:Control_Law_unequal_Vector_Relative_Degree}
Consider the multivariable system defined as \eqref{eq:system} and the time-varying optimization problem defined as \eqref{eq:Time_Varing_Optimization_Problem}. Suppose that both Assumption \ref{ass:convexity} and Assumption \ref{ass:non-Unifrom_Vector_Relative_Degree} are satisfied, 
then the system will globally exponentially converge to the optimal solution of \eqref{eq:Time_Varing_Optimization_Problem}, by using  the control law: 
\begin{align}
    \mathbf{u}\!=\!\alpha(\mathbf{x},\boldsymbol{\zeta})\! +\!\beta(\mathbf{x},\boldsymbol{\zeta})\mathbf{R}^{-1}(\mathbf{x},\boldsymbol{\zeta})[\tilde{\alpha}(\boldsymbol{\xi}) \!+\!\tilde{\beta}(\boldsymbol{\xi})\mathbf{y}_{\rm imp}^{(k)}\!-\!\mathbf{p}(\mathbf{x},\boldsymbol{\zeta})]\label{eq:final_Control_Input_Unequal}
\end{align}where $\mathbf{y}_{\rm imp}^{(k)}$ be the solution of \eqref{eq:y_kth_derivative_solution}, the dynamic feedback function defined in \eqref{eq:Dynamic_Feedback_Function} and the auxiliary states $\xi$ satisfy \eqref{eq:unequal_auxilliary_state_xi_dyanmic_equation}.
Moreover, the following inequalities hold:
\begin{align*}
    &\| \mathbf{y}(t)-\mathbf{y}^*(t)\|_2 \leq Ce^{-\alpha t}, \\
    &0\leq f_0(\mathbf{y}(t),t) - f_0(\mathbf{y}^*(t),t) \leq m_fC^2e^{-2\alpha t},\\
    & 0 < C = \left(\tfrac{c^2}{m_f^2} \sum_{j=0}^{k-1}\nolimits \norm{{\nabla_\mathbf{y}^{(j)} f_0(\mathbf{y}(0),0)}}_2^2)\right)^{\frac{1}{2}} < \infty,\nonumber
\end{align*}
for some constant $C > 0$, $  -\alpha = \max \Re(\lambda_i)+ \epsilon ,i \in [1...n],$ for some  $ \epsilon >0$ small enough.\\

\textit{Proof}: See Appendix \ref{prf:Control_Law_unequal_Vector_Relative_Degree}.
\end{thm}
}
{
\begin{ass}[Non-Uniform Vector Relative Degree]\label{ass:non_uniform_Vector_Relative_Degree_conference}
The multivariable nonlinear system \eqref{eq:composite-system} has vector relative degree $\{r_1,\dots,r_m\}$ and $r_1+r_2+\dots+r_m=n$.
\end{ass}
As a result of Assumption \ref{ass:non_uniform_Vector_Relative_Degree_conference}, the order of Lie differentiation of each channel is different (c.f.\eqref{eq:relative-degree-matrix}) and we cannot directly design the trajectory as in \eqref{eq:Unicycle_yddot}. However, remember that according to Lemma \ref{Lem:Dynamic_Feedback_Linearization}, the system is transformed into $\mathbf{\dot{z} = Az+Bv}$, with $(\mathbf{A,B})$ in Brunovsky Canonical Form. As a matter of fact, the input $v_i$ controls only the output $y_i$ throughout out a chain of $r_1$ integrators. If $\{r_1,...r_m\}$ are not equal, we can always introduce $k-r_i$ auxiliary states (integrators) for each channel $y_i$, where $k = \max \{ r_1,r_2, \dots ,r_m  \}$ and define the new input $s_i$ accordingly. Notice that this construction makes a dynamic extension of $\mathbf{\dot{z} = Az+Bv}$ that posses uniform order of Lie differentiation of each channel. For example, for channel $y_i$, we introduce the following states $
    \xi_1^i = v_i ,
    \xi_2^i = \dot{\xi}_1^i , 
    \dots,
    \dot{\xi}_{k-r_i}^i = s_i.$
 More specifically, the auxiliary states $\xi$ should satisfy the following dynamic:
\begin{subequations}\label{eq:unequal_auxilliary_state_xi_dyanmic_equation}
\begin{align}
    & \boldsymbol{ v} = \tilde{\alpha}(\boldsymbol{\xi}) + \tilde{\beta}(\boldsymbol{\xi})\boldsymbol{s},  \\ 
 &   \Dot{\boldsymbol{\xi}} =\tilde{ \gamma}(\boldsymbol{\xi}) +\tilde{ \delta}(\boldsymbol{\xi})\boldsymbol{s}.
\end{align}
\end{subequations}
Then the feedback function \eqref{eq:Dynamic_Feedback_Function}, the auxiliary states dynamic of $\boldsymbol{\xi}$ \eqref{eq:unequal_auxilliary_state_xi_dyanmic_equation}, and a state diffeomorphism $ \mathbf{z}= \Phi(\mathbf{x},\boldsymbol{\zeta,\xi})$ will transform the composite system \eqref{eq:composite-system} into $\dot{\mathbf{z}} = \mathbf{Az+Bs}$, with $\mathbf{A,B}$ in Brunovsky Canonical Form. 

By computing the higher derivatives of output channels, we can implicitly design the trajectory for $\mathbf{y}$ using $\mathrm{col}(\nabla_\mathbf{y} f_0(\mathbf{y},t),\dots,\nabla^{(k-1)}_\mathbf{y} f_0(\mathbf{y},t))$ as a proxy for optimality error, where the goal is to construct the following dynamical system:
\begin{align}
    \begin{bmatrix}
       \dot{\nabla}_\mathbf{y} f_0(\mathbf{y},t) \\ \vdots \\ \nabla^{(k)}_\mathbf{y} f_0(\mathbf{y},t)
    \end{bmatrix} = \mathbf{H}
    \begin{bmatrix}
       \nabla_\mathbf{y} f_0(\mathbf{y},t) \\ \vdots \\ \nabla^{(k-1)}_\mathbf{y} f_0(\mathbf{y},t)
    \end{bmatrix}   ,\label{eq:K-th_order_Optimality_Error}
\end{align} with
\begin{align*}
    \mathbf{H} = \begin{bmatrix*}
    0 &1 &0 &\dots &0 \\
    0 &0 &1 &\dots &0 \\
    \vdots & \vdots &  &\ddots & \vdots\\
    a_0 & a_1& a_2 &\dots &a_{k-1}
    \end{bmatrix*} \otimes \mathbf{I}_m \label{eq:Hurwitz_Matrix_Big_H}
\end{align*}{}

being Hurwitz.The following technical lemma will be used during the calculation of new optimality error state.

\begin{lem}[Gradient Time Differentiation]\label{lem:Diffrentiating_Gradient_K_Times}
Differentiating the gradient $\nabla_\mathbf{y} f_0(\mathbf{y},t)$ with respect to time $k-$times yields:
\begin{equation}
\begin{aligned}
    \nabla_\mathbf{y}^{(k)} f_0(\mathbf{y},t) &= \sum_{m = 0}^{k-1} \binom{k-1}{m} \nabla_{\mathbf{yy}}^{(m)}f_0(\mathbf{y},t)\mathbf{y}^{(k-m)}\\
    &+\nabla_{\mathbf{y}t}^{(k-1)}f_0(\mathbf{y},t),
\end{aligned} \label{eq:Differentiate_gradient_K_Times}
\end{equation}
where $\binom{k-1}{m}$ represent the binomial coefficients.

\textit{Proof}: See \cite{zheng2019implicit}.\\
\end{lem}
Combining \eqref{eq:K-th_order_Optimality_Error} and \eqref{eq:Differentiate_gradient_K_Times}, we can implicitly design the trajectory for $\mathbf{y}$ by:
\begin{align}
& \mathbf{y}_{\rm imp}^{(k)} = \nabla_{\mathbf{yy}}^{-1}f_0(\mathbf{y},t)[\sum_{i = 0} ^{k-1} a_i\nabla_{\mathbf{y}}^{(i)}f_0(\mathbf{y},t) -  \nabla_{\mathbf{y}t}^{(k-1)}f_0(\mathbf{y},t) \nonumber \\
& - \sum_{m=1}^{k-1}\binom{k-1}{m} \nabla_{\mathbf{yy}}^{(m)}f_0(\mathbf{y},t)\mathbf{y}^{(k-m)}  ] \label{eq:y_kth_derivative_solution}.
\end{align}
Now, we formally provide our solution for systems with non-uniform relative degree.
\begin{thm}[Control Law for General Vector Relative Degree System]\label{thm:Control_Law_unequal_Vector_Relative_Degree}
Consider the multivariable system defined as \eqref{eq:system} and the time-varying optimization problem defined as \eqref{eq:Time_Varing_Optimization_Problem}. Suppose that both Assumption \ref{ass:convexity} and Assumption \ref{ass:non_uniform_Vector_Relative_Degree_conference} are satisfied, 
then the system will globally exponentially converge to the optimal solution of \eqref{eq:Time_Varing_Optimization_Problem}, by using  the control law: 
\begin{align*}
    \mathbf{u}\!=\!\alpha(\mathbf{x},\boldsymbol{\zeta})\! +\!\beta(\mathbf{x},\boldsymbol{\zeta})\mathbf{R}^{-1}(\mathbf{x},\boldsymbol{\zeta})[\tilde{\alpha}(\boldsymbol{\xi}) \!+\!\tilde{\beta}(\boldsymbol{\xi})\mathbf{y}_{\rm imp}^{(k)}\!-\!\mathbf{p}(\mathbf{x},\boldsymbol{\zeta})]
    \label{eq:final_Control_Input_Unequal},
\end{align*}where $\mathbf{y}_{\rm imp}^{(k)}$ is given in \eqref{eq:y_kth_derivative_solution}, the dynamic feedback function defined in \eqref{eq:Dynamic_Feedback_Function} and the auxiliary states $\xi$ satisfy \eqref{eq:unequal_auxilliary_state_xi_dyanmic_equation}.
More specifically, the following inequalities hold:
\begin{align*}
    &\| \mathbf{y}(t)-\mathbf{y}^*(t)\|_2 \leq Ce^{-\alpha t}, \\
    &0\leq f_0(\mathbf{y}(t),t) - f_0(\mathbf{y}^*(t),t) \leq m_fC^2e^{-2\alpha t},\\
    & 0 < C = \left(\tfrac{c^2}{m_f^2} \sum_{j=0}^{k-1}\nolimits \norm{{\nabla_\mathbf{y}^{(j)} f_0(\mathbf{y}(0),0)}}_2^2)\right)^{\frac{1}{2}} < \infty,\nonumber
\end{align*}
for some constant $C > 0$, $  -\alpha = \max \Re(\lambda_i)+ \epsilon ,i \in [1...n],$ for some  $ \epsilon >0$ small enough.\\

\textit{Proof}: See \cite{zheng2019implicit}.
\end{thm}
}

\section{Numerical Examples}
In this section, we illustrate how to leverage the time-varying optimization algorithm to solve the following robot tracking problems.
\subsection{Switching Tracking Goals}\label{sec:simulation:1}
 Consider a wheeled mobile robot \eqref{eq:Unicycle_System_Dynamic} charged with the task of tracking two moving objects sequentially. In the first time interval $[t_0, t_s]$, the agent is required to track the first object and in the second time interval $[t_s,t_f]$ gradually switched to track the second object. The equivalent time-varying optimization problem takes the following form:
\begin{align*}
    \min_{\mathbf{y}} S(t)\| \mathbf{y}-\mathbf{y}_1^d(t)\|^2_2+ (1-S(t))\| \mathbf{y}-\mathbf{y}_2^d(t)\|^2_2,
\end{align*}where $\mathbf{y}(t)$ is the robot position satisfying \eqref{eq:Unicycle_System_Dynamic}, $\mathbf{y}_1^d(t), \mathbf{y}_2^d(t)$ represents the position of moving objects at time $t$ respectively. 
The smooth switch function $S(t)$ takes the form: $S(t) = 0.5-0.5\tanh(\frac{t-a}{b})$, where the parameters $a$ and $b$ can be used to define the switch point and the smoothing level. The target trajectories are designed via time parametric representation, where we use differential flatness in this trajectory generation problem \cite{martin2003flat}.
\ifthenelse{\boolean{arxiv}}{
Specifically, we parametrize the components of the flat output $\boldsymbol{\phi}_1 =\mathbf{y}= [x_1,x_2],\boldsymbol{\phi}_2 =\dot{\mathbf{y}}$, by \begin{align*}
    \phi_i(t) = \sum_{j = 0}^{n-1} A_{ij}\lambda_j(t),
\end{align*} where the $\lambda_j(t) = t^j$ are the standard polynomial basis functions and the degree of the polynomial is set to be $n = 4$. Thus, the trajectory generation problem reduces from finding a function to finding a set of parameters. }
{}
\begin{figure}[h]
\centering
\includegraphics[width=0.45\textwidth]{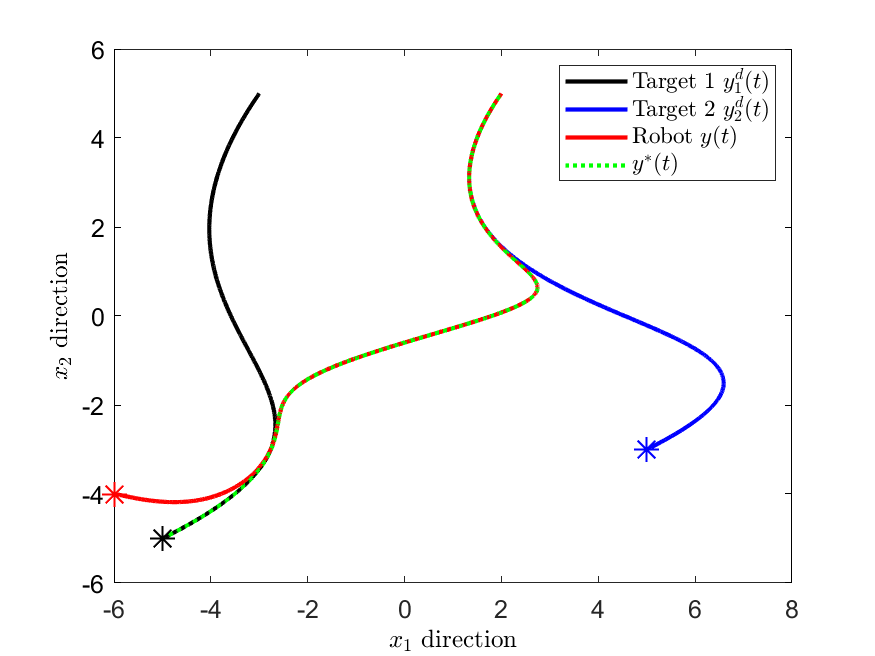}
    \caption{Trajectory of the optimal solution $\mathbf{y}^*(t)$ (dashed green), the robot (solid red) and the objects (black for 1 and blue for 2). The robot converge to the target trajectory, which is to track the first object from $[0s,5s]$ and gradually switch to track the second object in $[5s,15s]. $}
    \label{fig:2}
    \end{figure}
    
The resulting trajectories we proposed are illustrated in Figure \ref{fig:2}, with $a \!= \!10, b\!= \!1.5$. It can be observed that the robot successfully tracks the first object up to time $t_s\!=\!5s$, gradually switching to the second object until $t_f \!= \!15s$, and track the second object until simulation stops. Particularly, the randomly picked starting positions (highlighted using asterisk) for the two objects are $[-5,-5]$ and $[5,-3]$ and the agent is positioned randomly near the starting position, which is $[-5,4]$. We set $t_0 = 0s$ and the total simulation time is $ 20s$. 
For this implementation, the differential equation \eqref{eq:Unicycle_System_Dynamic} is solved based on an explicit Runge-Kutta $(4,5)$ formula, the Dormand-Prince pair.

\subsection{Multi-robot Navigation}\label{sec:simulation:2}
In this numerical example, two agents are required to track two moving objects respectively, but the maximum distance between two agents is limited (e.g., due to communication or formation constraints). We assume $\mathbf{y}_1(t), \mathbf{y}_2(t)$ representing the current position of each robot, whose dynamic are unicycles satisfying \eqref{eq:Unicycle_System_Dynamic}. We consider the following time-varying optimization problem for this task:
\begin{align*}
    \min_{\mathbf{y}_1,\mathbf{y}_2} \| \mathbf{y}_1\!-\!\mathbf{y}_1^d(t)\|^2_2\!+\! \| \mathbf{y}_2\!-\!\mathbf{y}_2^d(t)\|_2^2 
    \!+\! H(\| \mathbf{y}_1\!-\!\mathbf{y}_2\|_2^2),
\end{align*} where $\mathbf{y}_1^d(t), \mathbf{y}_2^d(t)$ represents the current position of the moving object. $H(x) = \alpha \tan(\frac{x\pi}{2d})^2$ is a smooth barrier function, where the parameter $d$ determines the maximum distance allowed for the two agents and $\alpha$ determines the flatness of penalty gain. In this scenario, although our theory do not exactly holds, since the barrier barrier function is not defined globally, as long as the initial conditions are not violated, the numerical result suggests that our algorithm can be applied beyond the presented assumptions.

\begin{figure}[h]
\centering
    \includegraphics[width=0.45\textwidth]{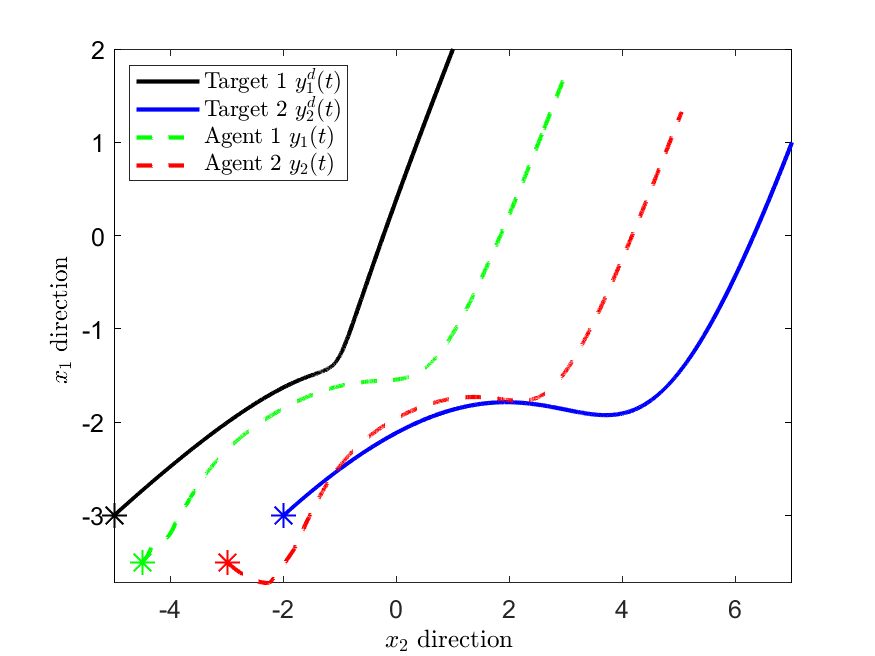}
    \caption{Trajectories of two objects $\mathbf{y}_1^d(t), \mathbf{y}_2^d(t)$ (solid) and two agents $\mathbf{y}_1, \mathbf{y}_2$ (dashed). Agents succeed in tracking objects while satisfying distance constraint between them. }
    \label{fig:3}
    \end{figure}
The trajectories for the objects were also in time parametric representation, following the same computing procedure as in the previous section. Particularly, the randomly picked starting position (using asterisk) for two objects are $[-5,-3]$ and $[-2,-3]$, respectively. The maximum allowed distance is set to be $d = 2$, and the gain is $\alpha = 1e-8$. As to the agents, they are positioned randomly near the starting position, while satisfying the distance constraint between them, which are $[-4.5,-3.5]$ and $[-3.5,-3.5]$ (using asterisk). For this implementation, the differential equation \eqref{eq:Unicycle_System_Dynamic} is solved using the same procedure as in Section \ref{sec:simulation:1}.
The resulting trajectories are illustrated in Figure \ref{fig:3}, where both robots, starting from arbitrary positions succeed in tracking the moving object and keep the maximum distance within limits simultaneously.

\section{Conclusion}
 In this paper we develop an optimization-based framework for joint real-time trajectory planning and feedback control of feedback-linearizable systems. We implicitly define a target trajectory as the optimal solution of a time-varying optimization problem, which is strongly convex and smooth. For systems that are (dynamic) full-state linearizable, the proposed control law transforms the nonlinear system into an optimization algorithm of sufficiently high order. Under reasonable assumptions, our method globally asymptotically converges to the time-varying optimal solution of the original problem. 
%
Further work include: (i) adding equality and inequality time-varying constraints in the framework and (ii) considering more general nonlinear system that are not feedback linearizable.

\ifthenelse{\boolean{arxiv}}{
\section{Acknowledgements}
The authors would like thank Prof. George Pappas for motivating this work, and Prof. Marin Kobilarov and Mr. Rajni Bansal for their involvement in a class project that initiated this research.
}
{}
\ifthenelse{\boolean{arxiv}}{ 
\appendix

\subsection{Proof of Lemma \ref{lem:Diffrentiating_Gradient_K_Times}}\label{Prf:Diffrentiating_Gradient_K_Times}
We prove by mathematical induction. First we consider when $k = 1$ and $2$.
\begin{align*}
    \dot{\nabla}_\mathbf{y} f_0(\mathbf{y},t) &= \frac{\partial \nabla_\mathbf{y} f_0(\mathbf{y},t) }{\partial \mathbf{y}}\dot{\mathbf{y}}+\frac{\partial \nabla_\mathbf{y} f_0(\mathbf{y},t) }{\partial t}  \nonumber\\
    &= \nabla_{\mathbf{yy}}f_0(\mathbf{y},t)\dot{\mathbf{y}}+\nabla_{\mathbf{y}t}f_0(\mathbf{y},t)\\
    \ddot{\nabla}_\mathbf{y} f_0(\mathbf{y},t) &= \frac{d}{dt}(\nabla_{\mathbf{yy}}f_0(\mathbf{y},t)\dot{\mathbf{y}}+\nabla_{\mathbf{y}t}f_0(\mathbf{y},t)) \nonumber\\
    &= \nabla_{\mathbf{yy}}f_0(\mathbf{y},t)\ddot{\mathbf{y}}+\dot{\nabla}_{\mathbf{yy}}f_0(\mathbf{y},t)\dot{\mathbf{y}}+\dot{\nabla}_{\mathbf{y}t}f_0(\mathbf{y},t)
\end{align*} 
We want to show that for every $k \geq k_0$, $k_0 \geq 2$, if the statement holds for $k$, then it holds for $k+1$. 
\begin{align*}
    \nabla_\mathbf{y}^{(k)} f_0(\mathbf{y},t) &= \sum_{m = 0}^{k-1} \binom{k-1}{m} \nabla_{\mathbf{yy}}^{(m)}f_0(\mathbf{y},t)\mathbf{y}^{(k-m)} \nonumber\\
    &+\nabla_{\mathbf{y}t}^{(k-1)}f_0(\mathbf{y},t)
\end{align*}

Using the binomial theorem we obtain:
\begin{align*}
    \nabla_\mathbf{y}^{(k+1)} f_0(\mathbf{y},t) &= \frac{d}{dt}(\sum_{m = 0}^{k-1} \binom{k-1}{m} \nabla_{\mathbf{yy}}^{(m)}f_0(\mathbf{y},t)\mathbf{y}^{(k-m)}) \nonumber\\
    &+\frac{d}{dt}(\nabla_{\mathbf{y}t}^{(k-1)}f_0(\mathbf{y},t)) \nonumber\\
    &= \sum_{m = 0}^{k} \binom{k}{m} \nabla_{\mathbf{yy}}^{(m)}f_0(\mathbf{y},t)\mathbf{y}^{(k+1-m)} \nonumber
    \\&+\nabla_{\mathbf{y}t}^{(k)}f_0(\mathbf{y},t),
\end{align*}which completes the proof.

\subsection{Proof of Theorem \ref{thm:Control_Law_uniform_Vector_Relative_Degree}} \label{prf:BCF_UniformRelativeDegree}

By uniformly strong convexity of $f_0(\mathbf{y},t)$ in $\mathbf{y}$, the Hessian inverse $\nabla_{\mathbf{yy}}^{-1}f_0(\mathbf{y},t)$ is defined for all $t \geq 0$. Because the vector relative degree of the nonlinear system is $r_1=\dots=r_m=k$, which means $\mathbf{y}^{(k)}$ has a linear relationship with new input $\mathbf{v}$. According to Lemma \ref{lem:Diffrentiating_Gradient_K_Times}, we have \eqref{eq:Differentiate_gradient_K_Times}.
Furthermore, as a result of Theorem \ref{Lem:Dynamic_Feedback_Linearization}, feedback function of the form \eqref{eq:final_Control_Input} results in $\mathbf{y}^{(k)}=  \mathbf{y}_{\rm imp}^{(k)} $, where $\mathbf{y}_{\rm imp}^{(k)}$ is the solution of \eqref{eq:y_kth_derivative_solution}.

Now, we are able to construct the desired dynamical system \eqref{eq:K-th_order_Optimality_Error}, where $\mathbf{H}$ is the designed Hurwitz matrix, and the solution of this ODE is:
\begin{align}
    \begin{bmatrix}
       \nabla_\mathbf{y} f_0(\mathbf{y},t) \\ \vdots \\ \nabla^{(k-1)}_\mathbf{y} f_0(\mathbf{y},t)
    \end{bmatrix} &= e^{\mathbf{H}t}
    \begin{bmatrix}
       \nabla_\mathbf{y} f_0(\mathbf{y}(0),0) \\ \vdots \\ \nabla^{(k-1)}_\mathbf{y} f_0(\mathbf{y}(0),0)
    \end{bmatrix}
\end{align}
where $\mathbf{y}(0) \in R^m$ is the initial point. By taking the Frobenius norms of both sides and applying Theorem \ref{thm:exp-convergence} we obtain
\begin{align}
\sum_{j=0}^{k-1} \norm{{\nabla_\mathbf{y}^{(j)} f_0(\mathbf{y},t)}}_2^2
    \leq c^2e^{-2\alpha t}( \sum_{j=0}^{k-1} \norm{{\nabla_\mathbf{y}^{(j)} f_0(\mathbf{y}(0),0)}}_2^2) \label{eq:Left_Norm_Inequality}
\end{align}
for some constant $c>0$, $  -\alpha = \max \Re(\lambda_i)+ \epsilon ,i \in [1...n],$ for some  $ \epsilon >0$ small enough.

Next, we use the mean-value theorem to expand $\nabla _\mathbf{y} f_0(\mathbf{y},t)$ with respect to $\mathbf{y}$ as follows, where $\boldsymbol{\eta(t)}$ is a convex combination of $\mathbf{y}(t)$ and $\mathbf{y}^*(t)$. Additionally using the fact that $\nabla _\mathbf{y} f_0(\mathbf{y}^*(t),t)=0$ for all $t \geq 0$, we obtain:
\begin{align}
    \mathbf{y}(t)-\mathbf{y}^*(t) = \nabla_{\mathbf{yy}}^{-1} f_0(\boldsymbol{\eta}(t),t)\nabla _\mathbf{y} f_0(\mathbf{y}(t),t).
\end{align}
It follows from Assumption \ref{ass:convexity}, that $\norm{\nabla_{\mathbf{yy}}^{-1}f_0(\mathbf{y},t)}_2 \leq m_f^{-1}$.
Taking the norm on both sides together with equation \eqref{eq:Left_Norm_Inequality} we have:
\begin{align}
     &\norm{{\mathbf{y}(t)-\mathbf{y}^*(t)}}_2  \leq C e^{-\alpha t}, \nonumber\\
    & 0 \leq C = \left(\tfrac{c^2}{m_f^2} \sum_{j=0}^{k-1}\nolimits \norm{{\nabla_\mathbf{y}^{(j)} f_0(\mathbf{y}(0),0)}}_2^2)\right)^{\frac{1}{2}} < \infty.
\end{align}\\
On the other hand, convexity of $f_0(\mathbf{y},t)$ implies that for each $t \geq 0$
\begin{align}
    0 \!\leq\! f_0(\mathbf{y},t)\! - \!f_0(\mathbf{y}^*,t) \!\leq\! \nabla _\mathbf{y} f_0(\mathbf{y},t)^T (\mathbf{y}\!-\!\mathbf{y}^*)
\end{align} By applying Cauchy-Swhartz inequality on the right hand side we obtain;
\begin{align}
 0 \leq {f_0(\mathbf{y}(t),t) - f_0(\mathbf{y}^*(t),t)} \leq m_fC^2e^{-2\alpha t}
\end{align}
which completes the proof.

\subsection{Proof of Theorem \ref{thm:Control_Law_unequal_Vector_Relative_Degree}} \label{prf:Control_Law_unequal_Vector_Relative_Degree}
Feedback function of the form \eqref{eq:final_Control_Input_Unequal} results in $\mathrm{col}(y_1^{(k)},\dots ,y_m^{(k)})
=  \mathbf{y}_{\rm imp}^{(k)} $, where $\mathbf{y}_{\rm imp}^{(k)}$ is the solution of \eqref{eq:y_kth_derivative_solution}. Rest of the proof follows \ref{prf:BCF_UniformRelativeDegree}.
}
{}
\bibliographystyle{IEEEtran}

\bibliography{main.bib}
\end{document}